# DO RADIOACTIVE HALF-LIVES VARY WITH THE EARTH-TO-SUN DISTANCE?

J.C. Hardy*, J.R. Goodwin and V.E. Iacob[#]

*Cyclotron Institute, Texas A&M University, College Station, TX 77845-3366, USA*

**Abstract**

Recently, Jenkins, Fischbach and collaborators have claimed evidence that radioactive half-lives vary systematically over a ±0.1% range as a function of the oscillating distance between the Earth and the Sun, based on multi-year activity measurements. We have avoided the time-dependent instabilities to which such measurements are susceptible by directly measuring the half-life of $^{198}$Au ($t_{1/2}$ = 2.695 d) on seven occasions spread out in time to cover the complete range of Earth-Sun distances. We observe no systematic oscillations in half-life and can set an upper limit on their amplitude of ±0.02%.

## 1. Introduction

In a series of papers, which began in 2009 and still continues, Jenkins, Fischbach and collaborators claim that radioactive half-lives vary as a function of the distance between the Earth and the Sun at the time of measurement (see, for example, Jenkins *et al.*, 2010). These claims are principally based on their interpretation of two sets of data taken some time ago by others, at the Brookhaven National Laboratory (Alburger *et al.*, 1986) and at the Physikalisch Technische Bundesanstalt (Siegert *et al.*, 1998). The BNL measurements compared the decay rate of $^{32}$Si ($t_{1/2}$ = 172 yr) to that of $^{36}$Cl ($t_{1/2}$ = 3.0×10$^5$ yr) on a regular basis over four years; the authors used an end-window gas-flow proportional counter to detect decay β particles. The PTB measurements were of the decay rate of $^{226}$Ra ($t_{1/2}$ = 1600 yr) and extended over an 11-year period; in this case, a high-pressure 4πγ ionization chamber was used. The data from both groups show a weak but statistically significant oscillatory behavior of decay rate with a one-year period, the highest rate (shortest half-life) occurring a few days after the Earth-Sun orbit had reached its perihelion.

At the time of publication, both groups acknowledged the oscillations in their data, with BNL noting that it corresponded with seasonal variations in temperature and humidity, which could have affected the relative absorption of the β particles from $^{32}$Si and $^{36}$Cl, while PTB attributed it to background radioactivity such as radon and its daughter products, which are known to show seasonal concentration changes. More recently, Schrader (2010) has pointed out that the variations in the PTB results disappear or completely change their structure when a different current-measurement technique is used. Nevertheless, Jenkins *et al.* (2010) have taken both data sets at face value and proposed much more fundamental causes, such as possible changes in the magnitudes of fundamental constants – the fine structure constant or the electron-to-proton mass ratio – or changes in the flux of solar neutrinos (see Fischbach *et al.*, 2009). More recently, Fischbach *et al.* (2011) even speculate that new objects they call "neutrellos" could be responsible.

Only two recent publications have presented experimental results that contradict temporal oscillations in decay rates. One, by Cooper (2009), uses data from the Cassini spacecraft to show that the α-decay rate of $^{238}$Pu ($t_{1/2}$ = 87.7 yr), which is used to power the craft, shows no change correlated with its distance from the Sun. The second, by Norman *et al.* (2009), analyzes previously published data on long-lived α, β$^+$, β$^-$ and electron-capture emitters and finds no evidence of correlations between the decay rates and the Earth-Sun distance.

So far, all the experimental data used in this controversy – the BNL and PTB results interpreted so significantly by Jenkins *et al.* (2010), and the contrary results that appear in the papers by Cooper (2009) and Norman *et al.* (2009) – have been from successive activity measurements of long-lived radionuclides. As already mentioned, such measurements extending over months or years are susceptible to systematic effects and instabilities arising from changes in temperature, humidity, background radiation and instrumental drifts. To avoid these problems, we have followed a quite different approach, making seven individual half-life measurements of a shorter-lived radionuclide, $^{198}$Au ($t_{1/2}$ = 2.7 d), spread out in time so that all seven measurements together span the full range of Earth-Sun distances. By depending on direct, relatively short half-life measurements, rather than separate activity measurements spaced over a long time, we substantially reduce the effects of environmental and instrumental variations.

## 2. Half-life measurements and analysis

When the first of the claims by Jenkins *et al.* (2010) appeared as an arXiv preprint, we had already made three sequential measurements of the half-life of the β decay of $^{198}$Au for another purpose, so we decided to make four more half-life measurements to ensure that an approximately evenly-spaced set of Earth-Sun distances were being sampled. Our measurement techniques have been presented previously (see Goodwin *et al.*, 2007 and Hardy *et al.*, 2010) so they will only be briefly described here.

For each half-life measurement we prepared a fresh source by activating a 99.99+% pure gold foil, 10 mm in diameter and 0.1 mm thick, for 10 seconds in a flux of $10^{10}$ neutrons/(cm$^2$ s) in the Texas A&M Triga reactor. The activated foil was then fastened securely inside a chamber, outside of which a 70% HPGe detector was placed facing the source through 3.5 mm of stainless steel. The total distance between the detector face and the activated gold sample was 45 mm, and this arrangement remained fixed throughout each complete measurement, in which the decay was recorded for more than ten half-lives.

Our half-life results were obtained from analysis of the decay of the 412-keV delayed γ-ray (in $^{198}$Hg), which follows the $\beta^-$ decay of $^{198}$Au. Signals from the HPGe detector were amplified and sent to an analog-to-digital converter, which was an Ortec TRUMP$^{TM}$-8k/2k card controlled by Maestro software, installed on a PC operating under Windows-XP. Six-hour γ-ray spectra from the HPGe detector were acquired sequentially for more than a month from each source. At least 100 γ-ray spectra were recorded for each measurement, with a total of 786 spectra being obtained for all seven measurements.

The TRUMP$^{TM}$ card uses the Gedcke-Hale method (Jenkins *et al.*, 1981) to correct for dead-time losses. By keeping our system dead time below 3% at all times, and recording all spectra for an identical pre-set live time, we ensured that our results were nearly independent of dead-time losses. However, to arrive at our ultimate precision we made a further small correction for residual rate-dependent losses – from pile-up for

example – which we determined experimentally to be $(5.5 \pm 2.5) \times 10^{-4}$ per 1% increase in dead time. Goodwin *et al.* (2007) give further details about how this correction was obtained.

The 412-keV γ-ray peak in each recorded spectrum was analyzed with GF3, a least-squares peak-fitting program in the RADware series (Radford, 2009). This program allows us to be very specific in determining the correct background for a particular peak, and the 412-keV peak in each spectrum was visually inspected to this end. So far as possible, the same criteria were applied to each recorded spectrum. The peak areas thus derived were corrected for residual rate-dependent losses as described above, and their decay with time was then analyzed by a maximum-likelihood fit to a single exponential. We have tested and verified the code we used, which is based on ROOT (Brun *et al.*, 1997), to 0.01% precision with Monte Carlo generated data. As a typical decay curve and fit, we show the results from our fifth measurement in Figure 1. Other decay curves from the first three measurements in this series have already appeared in previous publications (Goodwin *et al.*, 2007; Goodwin *et al.*, 2010).

## 3.    Results

The results from all seven measurements of the $^{198}$Au half-life are presented in Table I. The uncertainties quoted there are purely statistical and do not include provision for the uncertainty in the correction for residual losses, since that correction is correlated for all seven measurements and does not contribute to the differences among them. Each measurement is characterized by a date and a time, which is taken to be one mean-life of $^{198}$Au ($\tau = t_{1/2} / \ln 2 = 3.89$ d) after the actual start time of the measurement. The table also shows how the time of each measurement relates to the time of the preceding perihelion of the Earth-Sun system. Clearly there is no dependence of the half-life on this parameter within our limits of uncertainty. In fact, the weighted average of the seven measured half-lives is 2.69445(20) d with a normalized $\chi^2$ of 0.74 and a confidence level of 64%.

How does our result compare with the BNL-observed activity oscillations, upon which Jenkins, Fischbach *et al.* base their claims? In Figure 2 we present the BNL activity results (Alburger *et al.*, 1986) as displayed by Jenkins *et al.* (2010): the activity ratios $^{32}$Si/$^{36}$Cl are normalized to their average and appear as grey circles with error bars, which are plotted against the dates of their measurement over a period of four years from early 1982 to early 1986. The dotted curve shows the $1/R^2$ behavior of the Earth-Sun distance over the same period, which is very nearly a pure sine wave. We have therefore fitted the BNL data with a sine wave, which has a fixed one-year period but variable amplitude and phase shift. A least squares fit to the data, shown as the dashed curve in the figure, yielded an amplitude of $7.9(3) \times 10^{-4}$ and a phase shift of 35(2) d relative to the $1/R^2$ plot. This phase shift was also noted by Jenkins *et al.* (2010).

Since the BNL measurements were of activity not half-life we show in the final column of Table 1 our half-life results for $^{198}$Au expressed as decay rates, $\lambda$ (= ln2 / $t_{1/2}$), and we plot these values normalized to their average in Fig. 2 as black circles with error bars. The time scale for our measurements has been displaced by exactly 25 years from the BNL scale, so our data appear with the same perihelion synchronization. The horizontal shaded band shows the one-standard-deviation uncertainty limits on the average value from our results. As already noted, our data are statistically consistent with a constant half-life value to within a relative precision of $\pm 7 \times 10^{-5}$, an order of magnitude smaller than the amplitude of the oscillations attributed to the BNL data.

It is also of interest to compare our results for the $^{198}$Au half life with previously published values for the same quantity. However, as already noted, the uncertainties associated with $t_{1/2}$ in Table 1 are purely statistical. Although the correction for residual losses has been incorporated in the half-life value itself, the uncertainty in the correction, being common to all, has not been applied to the uncertainties in the table. If we now apply that common uncertainty to the average value of the $^{198}$Au half-life, we obtain the final result:

$$t_{1/2}(^{198}\text{Au}) = 2.69445 \pm 20 \text{ (stat.)} \pm 25 \text{(syst.) d.}$$

This agrees completely with 2.6950(3) d, the weighted average of all thirteen previously published measurements that claimed better than 0.15% precision. (For a listing of the previous measurements, see Goodwin *et al.*, 2007.) It also agrees with, but is more precise than, 2.6950(7) d, the value recommended in the most recent IAEA data evaluation (IAEA, 2007).

## 4. Conclusions

We have measured the half-life of $^{198}$Au on seven different occasions that spanned the full range of Earth-Sun distances from perihelion to aphelion. Each measurement covered ten half-lives of the decay and established the half-life with high precision. The results were consistent with one another to a relative precision of $\pm 7 \times 10^{-5}$, in contradiction with previous claims (*e.g.* Jenkins *et al.*, 2010) that half-lives changed by as much as $\pm 7.9 \times 10^{-4}$ over each cycle of the Earth-Sun distance. Our average half-life result also agrees well with the average of (extensive) world data.

Ockham's razor – of several competing theories, the simplest one is to be preferred – could be invoked to argue that we need go no farther than to demonstrate that a simple constant fits our data. However, we have taken another step to least-squares fit our data using the same sine wave with a one-year period as we used for the BNL data. Fixing the phase shift at 35 d, the value obtained from the BNL data, we obtained an amplitude of $1.7(5) \times 10^{-4}$ with a normalized $\chi^2$ of 0.22. If we let both the amplitude and the phase shift vary, the amplitude remains about the same but the phase shift goes to zero and the normalized $\chi^2$ becomes an unreasonable 0.07. Clearly there is no statistically significant support for the interpretation that Jenkins, Fischbach *et al.* have applied to the BNL data. Our results demonstrate that the $^{198}$Au half-life is independent of the Earth-Sun distance to within $\pm 7 \times 10^{-3}$ % and, even if one postulates the possibility of a half-life dependence on that distance, the upper limit on its amplitude is $2 \times 10^{-2}$ %. In either case, these limits are well below the $8 \times 10^{-2}$ % number obtained from the BNL data on $^{32}$Si. Whatever caused the oscillation in the BNL data, it cannot have been principally due to oscillations in the Earth-Sun distance.

**Acknowledgements**

The work of the Texas A&M authors is supported by the U.S. Department of Energy under Grant No. DE-FG02-93ER40773 and by the Robert A. Welch Foundation under Grant No. A-1397.

**Footnotes**

* Corresponding author. Tel: +01-979-845-1411; fax: +01-979-845-1899.

  E-mail address: hardy@comp.tamu.edu

# On leave from the National Institute for Physics and Nuclear Engineering, "Horia Hulubei," Bucharest, Romania.

**Tables:**

Table 1: Dates of, and results from, our seven measurements of the half-life of $^{198}$Au. The date in each case is taken to be the starting time of our measurement plus one mean-life of $^{198}$Au (3.89 d). The decay rate, $\lambda$, is related to the half-life, $t_{1/2}$, by $\lambda = \ln2 / t_{1/2}$

| No. | Date (year/month/day/hour) | Days since last perihelion | $t_{1/2}$ (d) | $\lambda$ (d$^{-1}$) |
|---|---|---|---|---|
| 1 | 2007/06/24/1.2 | 171.2 | 2.6953(6) | 0.25717(6) |
| 2 | 2007/08/13/19.2 | 222.0 | 2.6949(5) | 0.25721(5) |
| 3 | 2009/06/14/21.2 | 161.3 | 2.6948(9) | 0.25722(9) |
| 4 | 2009/09/25/19.9 | 264.2 | 2.6944(5) | 0.25725(5) |
| 5 | 2009/11/08/17.6 | 308.1 | 2.6942(4) | 0.25727(4) |
| 6 | 2010/01/10/17.2 | 7.7 | 2.6940(5) | 0.25729(5) |
| 7 | 2010/03/07/16.6 | 63.7 | 2.6942(5) | 0.25727(5) |

**Figures**

Figure 1: Decay of $^{198}$Au as obtained in our measurement #5. Experimental data appear as dots with error bars; the straight line is a fit to these data. Normalized residuals appear at the bottom of the figure. The dashed lines in the plot of residuals represent ±1 standard deviation from the fitted value.

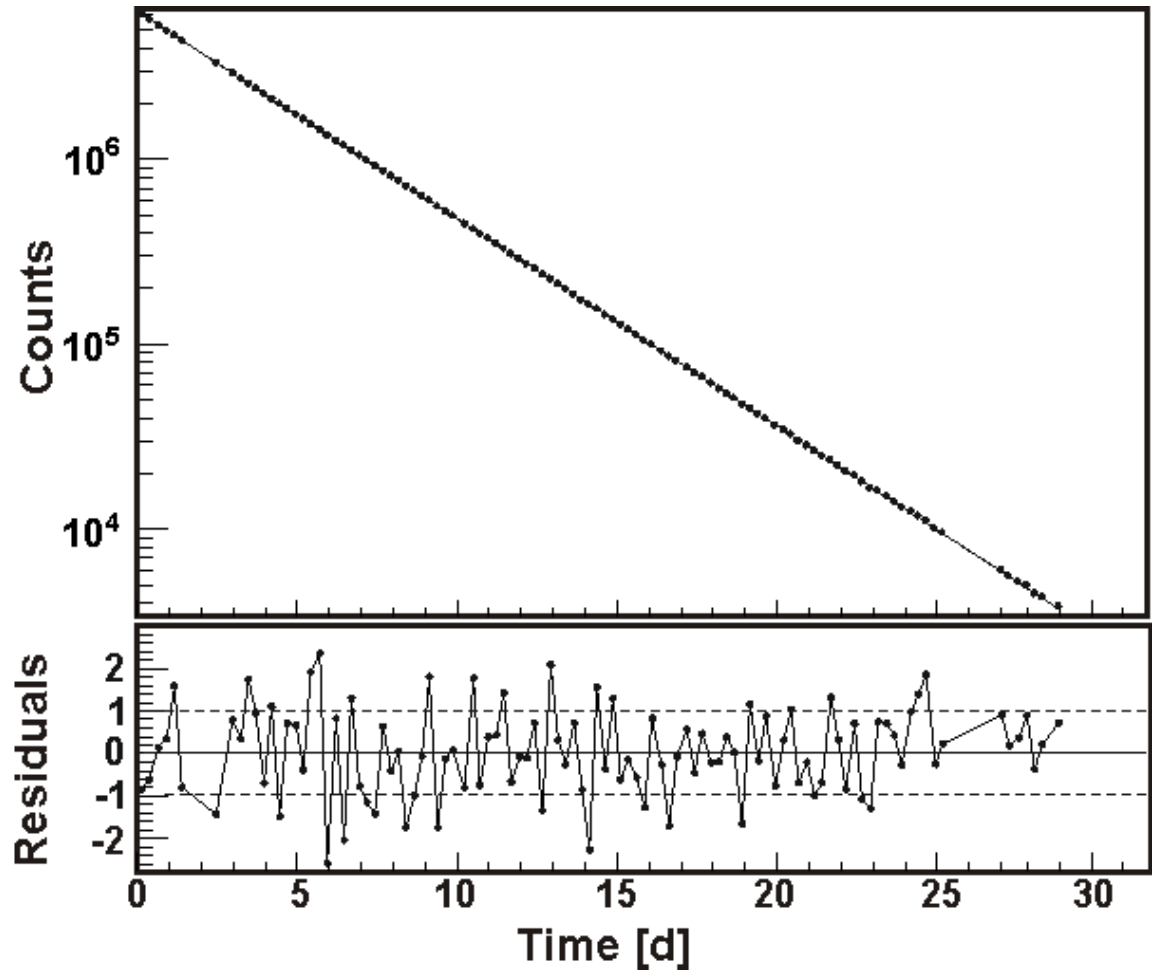

Figure 2: The BNL data for the activity ratio, $^{32}$Si/$^{36}$Cl, (Alburger *et al.*, 1986), as published by Jenkins *et al.* (2010), are plotted as grey circles with error bars (referred to the vertical axis at the left) against the dates of their measurement between 1982 and 1985 (horizontal axis at the bottom). The dotted curve shows the $1/R^2$ behavior of the Earth-Sun distance, where R is measured in astronomical units, a.u. (vertical scale at the right), over the same period; and the dashed curve gives our fit to the BNL data (see text). Our seven results for the decay rate of $^{198}$Au normalized to their average value (with the same vertical scale as the BNL data) are plotted as black circles with error bars against their dates (shown on the horizontal axis at the top, which is shifted exactly 25 years compared to the bottom scale).

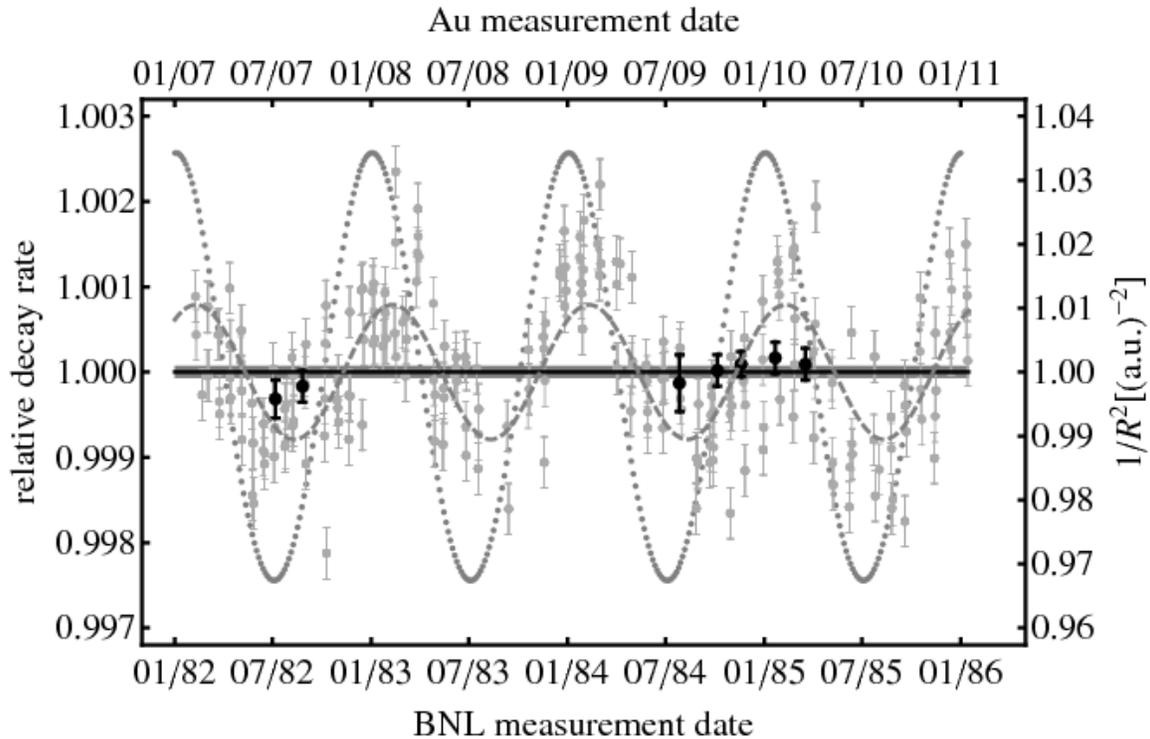